\documentstyle{aipproc}

\def\Mesz{M\'esz\'aros }

\def\etal{et al$.$ }

\begin{document}

\title{Theoretical Models of \\ Gamma-Ray Bursts 
\footnote{Invited review to appear in Procs. 4th Huntsville Gamma-Ray Burst Symposium, 1997} }
\author{P. \Mesz
\footnote{also Center for Gravitational Physics and Geometry, Pennsylvania State University} }
\address{Department of Astronomy \& Astrophysics, 525 Davey Lab, \\
      Pennsylvania State University, University Park, PA 16802 }
       
\maketitle

\begin{abstract}
Models of gamma ray bursts are reviewed in the light of recent observations
of afterglows which point towards a cosmological origin. The physics of 
fireball shock models is discussed, with attention to the type of
light histories and spectra during the gamma-ray phase.
The evolution of the remnants and their afterglows is considered, as well 
as their implications for our current understanding of the mechanisms
giving rise to the bursts.

\end{abstract}

\section*{Introduction}

The discovery of X-ray, optical and radio afterglows of gamma-ray bursts 
(GRB) amounts to a major qualitative leap in the type of independent 
observational handholds on these objects. Together with existing $\gamma$-ray 
signatures, these provide significantly more severe constraints on possible 
models, and may indeed represent the light at the end of the tunnel 
for understanding this long-standing puzzle of astrophysics.

The report of long wavelength observations of GRB 970228 over time scales 
of days to weeks at X-ray (X), and months at optical (O) wavelengths 
(Costa \etal, 1997) was the most dramatic recent development in the field. 
In this and subsequent IAU circulars, it was pointed out that the overall 
behavior of the long term radiation agreed with theoretical expectations 
from the simplest relativistic fireball afterglow models published in 
advance of the observations (\Mesz \& Rees, 1997a; see also Vietri, 1997a).
A number of theoretical papers were stimulated by this and subsequent 
observations (e.g. Tavani, 1997; Waxman, 1997a; Reichart, 1997; Wijers, \etal,
1997, among others),
and interest has continued to grow as new observations provided apparently 
controversial evidence for the distance scale, possible variability and
the candidate host (Sahu \etal, 1997). New evidence was added when
the optical counterpart to the second discovered afterglow (GRB 970508)
yielded a redshift lower limit placing it at a clearly cosmological
distance (Metzger \etal, 1997), and this was strengthened by the detection 
of a radio counterpart (Frail \etal, 1997; Taylor \etal, 1997) as well as 
evidence for the constancy of the associated diffuse source and
continued power law decay of the point source (Fruchter, \etal, 1997)

This new evidence reinforces the conclusions from previous work on the 
isotropy of the burst distribution which suggested a cosmological origin
(e.g Fishman \& Meegan, 1995). Observational material on this is provided 
chiefly by a superb data base (currently of over 1700 bursts in the 4B 
catalog) which 
continues being accumulated by the BATSE instrument, complemented by data 
from the OSSE and Comptel instruments on CGRO, as well as Ulysses, KONUS 
and other experiments. At gamma-ray energies, much new information has
been collected and analyzed, relevant to the spatial distribution, the
time histories, possible repeatability, spectra, and various types of
classifications and correlations have been investigated. At the same time, 
investigations of the physics of fireball models of GRB have continued to 
probe the $\gamma$-ray behavior of these objects, as well as the afterglows. 
Much of the recent theoretical work has concentrated on modeling the time 
structure expected from internal and external shock models, multi-wavelength 
spectra, the time evolution and the spectral-temporal correlations.

\section*{ The Dissipative Fireball Scenario }

The dissipative (or shock) fireball model is a fairly robust astrophysical 
scenario, independent of the particular type of progenitor, based only on 
the fact that it must inject the inferred large amount of energy inside 
the very small volume required by causality and the timescales 
characteristic of GRB (Rees \& \Mesz, 1992, \Mesz \& Rees, 1993). The 
observational and physical motivation for this generic scenario of GRB has 
been described in detail elsewhere (e.g. \Mesz, 1995). The very large energy
deposition  inside a small volume leads to characteristic photon energy 
densities which lead to an optically thick $\gamma e^\pm$ fireball that is 
highly super-Eddington. The resulting expansion must be highly relativistic
($\Gamma \sim 10^2-10^3$), in order to avoid having the observed 0.1-10 GeV 
photons degraded by photon-photon interactions, and to yield the right 
timescales and energies. The fireball initially is thermal, 
and converts most of its radiation energy into kinetic energy (bulk motion).
This kinetic energy of motion must be tapped via some dissipation mechanism,
which is most likely to be shocks, and these probably occur after the fireball 
becomes optically thin, as suggested by the nonthermal spectra.

The plasma, MHD and radiation physics involved in the fireball shock scenario
are familiar, being used in a number of other astrophysical situations. 
The basic ingredients, such as a high $\Gamma$ outflow, collisionless shocks,
magnetic field generation at some fraction of equipartition, acceleration
of electrons to a power law, efficient energy exchange between protons
and electrons, etc. are common features (or common problems, to varying 
degrees) in AGN, pulsars winds and supernova remnants. In AGN and possibly
pulsar winds, conditions qualitatively similar to those in GRB seem to 
obtain, and these sources are known to have in many cases efficiencies of at 
least tens of percent in converting bulk kinetic energy into nonthermal 
particles and radiation. As in those sources, for GRB fireballs it is assumed 
that the fluid approximation is valid whenever the usual plasma kinetic theory 
criteria are satisfied, e.g. that the dimensions of the region are much larger 
than the proton gyroradius or the proton Debye length. The shocks serve to 
reconvert the kinetic energy of the outflow into random energy, and to 
accelerate relativistic particles which can radiate a power law spectrum. 
The cosmological fireball shock model appears to have received strong 
confirmation from the afterglow observations, and from the fact that many
of the basic gamma-ray signatures can be understood within the framework 
of the model without undue parameter twisting.

The generic nature of this scenario stems from the fact that it is largely
{\it independendent} of the detailed nature of the primary energy  release
mechanism, whether it be a binary compact object merger (NS/NS or NS/BH,
e.g. Paczyn\'ski 1986), a ``failed Supernova Ib" (Woosley, 1992), a young
ultrastrongly magnetized pulsar (Usov, 1992), a ``hypernova" (Paczy\'nski,
1997), etc. This is because the primary mechanism is initially enshrouded 
in an optically thick pair fireball, which washes out most of the details,
the observed radiation being produced outside the pair photosphere. Some
information, however, may be carried through, e.g. in the details of the 
light curve (especially if this is due to internal shocks, see below).

A major theoretical question is how the very large bulk Lorentz factors 
inferred from observations are produced. Neutrino-antineutrino annihilation
leading to pairs (Eichler, \etal, 1989) have been proposed. Since the merger
would lead also to enormous radiation pressure, a baryon rich outflow
would however pollute the $e^\pm,\gamma$ fireball, but a clean fireball
might be achieved if tidal heating and annihilation occur before merger,
or if enough annihilations occur around the centrifugally evacuated binary 
rotation axis (\Mesz \& Rees, 1992). Numerical simulations using Newtonian
potentials (Ruffert \& Jahnka, 1997) indicate that this may not be 
straightforward, although effects like turbulent convection and magnetic 
fields could improve the pair luminosities. Matthews, \etal (1997) use
a general relativistic hydro code and conclude that both NS collapse to
black holes before merger, producing enough heating to power a pair
luminosity comparable to required estimates. The disagreement between
numerical simulation results is debated, and further refinements in models 
involving neutrino annihilation should be forthcoming. 

On the other hand, magnetic fields may be responsible for a large or even
dominant fraction of the relativistic stress tensor in the fireball.
Super-strong magnetic fields are probably generated during
the collapse of the rapidly rotating configuration (Usov, 1992, 
\Mesz \& Rees, 1992, Narayan, \etal, 1992, Thompson, 1994, Vietri, 1997a),
and this may contribute significantly to the energy density of the fireball.
Such fields could in fact be dynamically dominant around the rotation axis, 
especially if the central objects collapses to a black hole, leading to a 
Poynting dominated outflow which could be almost baryon-free (\Mesz \& Rees, 
1997b). Magnetic fields would, of course, also ensure a high radiation 
efficiency. MHD numerical simulations are difficult, as in pulsar winds
and AGN jets, and have not so far been done. In any case, it is worth 
stressing that the motivation for high bulk Lorentz factor ($\Gamma \gtrsim 
10^2$) outflows in GRB is largely observational, in particular the observation
of 0.1-10 GeV photons, which are hard to explain otherwise (e.g. Harding 
\& Baring, 1994).

\section*{ Gamma Ray Temporal Properties and Spectra }

Two types of fireball models have been discussed, both of which produce
the nonthermal spectrum via shocks occurring after the fireball has become
optically thin, as inferred from the nonthermal nature of the spectrum. 
These involve different explanations for the typical duration of the burst, 
and predict different time variabilities. In the first type (a) (e.g. 
\Mesz \& Rees, 1993) the shocks are those caused by interaction of the gaseous 
fireball ejecta with an external medium.  In this case the typical duration 
is given by the Doppler delayed arrival of the light from the beginning and 
end of the ejecta shell, or from the delay between surface elements within 
the light cone. This assumes that any ``intrinsic" burst duration is shorter 
than the above duration (impulsive approximation).  Any ``intrinsic" short 
time variability is washed out by the fact that radiation is received from a 
light cone and a finite width over which $\Gamma$ varies by at least a 
factor 2. (The afterglows discussed below are well fitted, in their 
overall average features, by the late stages of an external shock).

In the second type of model (b) (Rees \& \Mesz, 1994, Paczyn\'ski \& Xu, 
1994), the shocks leading to $\gamma$-rays are those which may be expected 
within the outflow itself, e.g. internal shocks caused by the catching up 
of faster portions with slower portions of the flow. These, if they occur, 
tend to do so at smaller radii than the previous external shocks, and the 
duration is likely to be given by the intrinsic duration of the
energy release (since the Doppler delayed light arrival or light cone
duration is likely to be shorter than the latter). One of the two stated
purposes for introducing this model is that it {\it does} specifically allow 
arbitrarily complicated light curves (Rees \& \Mesz, 1994), which are expected 
to reflect any ``intrinsic" variability injected at the base of the outflow.
These models are also referred to as wind models, or central engine models.

Detailed kinematical calculations (Fenimore, \etal, 1996) show explicitly 
some of the constraints imposed by observations on external shock light 
curves. Sari \& Piran (1997) showed analytically that external shocks in a 
blobby external medium would not be able to reproduce very complicated light 
curves with many subpulses, unless the efficiency is very low, $\lesssim 1\%$. 
Nonetheless, as shown by Panaitescu \& \Mesz (1997a), if magnetic
inhomogeneities are present or develop in the ejecta, and there is some
pre-beaming in the comoving frame, one can get up to 5-10 peaks with good
efficiency in an external shock light curve, and the spectral-temporal
correlations are close to the observed values. For bursts with a very large
number of subpulses, simulations of bolometric internal shock light curves 
(Daigne \& Mochkovich, 1997; Kobayashi, \etal, 1997) are in good qualitative 
agreement with the observations.

The nonthermal radiation spectrum of GRB is likely to be due to synchrotron 
or inverse Compton (IC) radiation of electrons or positrons accelerated in 
the optically thin shocks described above.  Particles accelerated, e.g. by 
a Fermi type mechanism, in the presence of modest magnetic fields lead to
nonthermal photon spectra similar to those observed (e.g. \Mesz, Rees \& 
Papathanassiou, 1994 for impulsive shock spectra, and Papathanassiou \& 
\Mesz, 1996 for wind spectra). Basically, two types of spectra are possible: 
those where the observed ``break" in the 50 KeV - 2 MeV range is due to the 
synchrotron
characteristic energy, or those where it is due to the IC upscattering of a
lower energy break (typically at optical energies) which itself is due to 
synchrotron. The latter requires smaller magnetic fields and smaller electron
minimum energies $\gamma_m$. Above $\gamma_m$ shock acceleration is assumed
to provide the electron power law responsible for the flattish $\nu F_\nu$
spectrum characteristic of bursts: an electron index $p \sim -3$ reproduces
this well. The burst spectra can satisfy the X-ray paucity condition (i.e.
the observation that generally $F_x \lesssim 3\times 10^{-2} F_\gamma$ during
the $\gamma$-ray burst), since below the break one expects a spectrum 
$\nu F_\nu \propto \nu^{4/3}$. On the other hand, spectra flatter than this 
can be easily obtained in an inhomogeneous magnetic field, or for a spatially 
varying bulk Lorentz factor, so that ``soft excess" bursts can also be
produced. In addition, one expects
significant simultaneous emission at GeV energies, and modest but detectable
simultaneous X-ray and optical emission (\Mesz \& Rees, 1993b, \Mesz, Rees 
\& Papathanassiou, 1994, Katz, 1994b, Papathanassiou \& \Mesz, 1996).
In addition, if GRB occur inside galaxies where the external medium has an
appreciable density, one would expect internal shock bursts to be followed
by external shock bursts, which can be relevant for, e.g. delayed GeV emission
(\Mesz \& Rees, 1994). In general one expects different spectral and temporal
properties for such compound bursts. A study of the properties of internal 
shocks followed by external shocks (Papathanassiou, 1997) provides constraints 
on the internal parameters of the outflow (duration, variability timescale and 
luminosity) in different external environments.

One of the signs of the development of the subject is that $\gamma$-ray
observations of GRB have become sufficiently detailed and extensive that 
they are beginning to probe questions of the internal physics of the 
models, such as the shock acceleration, the magnetic field equipartition
fraction, and the radiative efficiency involved.As far as the specific
radiation mechanism, Tavani (1996) has presented detailed synchrotron spectra 
calculated numerically with a distribution of shocked electrons produced by a
specific diffusive acceleration mechanism, and these were fitted to a variety 
of observed $\gamma$-ray spectra. Another investigation (Cohen, \etal, 1997)
aimed at testing the synchrotron hypothesis uses the fact that an
electron distribution with a low energy cutoff would produce a low
frequency asymptotic intensity spectrum with a slope of 1/3, while the
time integrated high frequency slope would be expected to be -1/2, 
compatible with a sample of BATSE spectra studied by these authors. 
This issue remains under discussion, e.g. Crider, \etal, 1997.
The problem of a relatively high radiative efficiency during the 
$\gamma$-ray event is, clearly, one of the requirements of a fireball shock,
or indeed of any other model. In particular, one needs to ensure that much 
of the energy carried in protons (if these are present and energetically 
dominant) is shared with the radiating electrons or pairs. Specific mechanisms 
have been proposed 
for this (Bykov \& \Mesz, 1996). A high radiative efficiency is natural 
in models where magnetic fields are prominent (e.g. Narayan, \etal, 1992,
Usov, 1994, Thompson, 1994). The electron-proton exchange would also be 
obviated in the reverse shock for scenarios involving Poynting dominated 
outflows where the inertia is mainly due to pairs (\Mesz \& Rees, 1997b), 
although in the late stages of deceleration the blast wave pushed ahead of
the ejecta will unavoidably include baryons. 

Another area of contact between models and observations 
is in the area of spectral-temporal correlations in the gamma-ray range.
Fenimore \& Sumner (1997) find that the observed spectral break energies 
decay in time faster than predicted from single shell analytic models. Crider, 
\etal 1997b argue that the evolution of the spectral break with integrated
photon flux may be a restriction on simple models. Numerical hydrodynamic 
simulations (Panaitescu \& \Mesz, 1997a) of external shock models indicate 
that many of the commonly observed correlations are well reproduced; among 
these are a brightness and hardness correlation, hardness and duration 
anti-correlation, a hard to soft evolution outside of intensity pulses, 
a break energy increasing with intensity during a pulse, the break
energy decreasing with increasing fluence, earlier pulses being harder, 
pulses being narrower at higher energies, etc. This is a continuing area 
of activity. 

\section*{ The Implications of Afterglows }

The breakthrough Beppo-SAX observation of GRB 970228 provided both a study 
of the long-term X-ray decay (extending over days) and 
an accurate localization permitting subsequent optical follow-ups (extending
to months). The X-ray and optical sources are both point-like, as expected
for a fireball at cosmological distances (dimension $\sim 0.1-1 $ pc after 
$\sim$ months). The spectra are nonthermal, as expected from the simplest 
model based on synchrotron radiation of shock-accelerated electrons from a 
decelerating shell interacting with an external medium (\Mesz \& Rees, 1997a), 
and as predicted, it decays as power law in time with an index close to the 
expected value (see also Vietri, 1997a; Waxman, 1997a; Reichart, 1997; Sari,
1997). Furthermore, a fuzzy extended source was identified
around the point source. While the initial optical magnitude of the point 
source decayed from $\sim 20$ to $\sim 24$, there was uncertainty  as to 
whether the diffuse source remained constant and whether it showed any 
proper motion (Caraveo, \etal, 1997).  However, the September 1997 HST images 
(Fruchter et al, 1997) have largely dispelled such doubts, indicating that
the diffuse source remained at $m_R \sim 25.5$ with negligible proper
motion, being compatible with a distant ($z \sim 1$), faint (possibly 
irregular) galaxy, while the optical point source is still present at 
$m_R \sim 28$, right along the extrapolation of the earlier power law.

A major highlight was the detection of the afterglow of GRB 970508,
which largely followed the pattern of GRB 970228, but which added 
considerable excitement because of new, even if not entirely unexpected, 
features. The most significant of these is that a redshift limit was 
obtained (Metzger, \etal, 1997) of $0.835 \leq z \lesssim 2.3$, based
on several systems of absorption lines. Another previously unobserved 
phenomenon was that the optical flux of the point source initially rose as a 
power law, followed by a decay similar to that of GRB 970228. This, in fact, 
is what one expects from a cloud where the spectrum has a peak initially 
above optical frequencies that shifts downwards during its expansion (Katz, 
1994b), and is in agreement with the \Mesz \& Rees (1997a) simplest model. 
Another previously unobserved feature was the detection of a radio afterglow
in GRB 970508, about a week after the outburst, peaking after weeks and then 
decaying slowly. With a self-absorption frequency around 5-10 GHz 
(overestimated in early calculations), this is also compatible with the 
'simplest' model (e.g. Waxman, 1997b, Katz \& Piran, 1997). Furthermore, 
scintillations in the radio spectrum, predicted by Goodman (1997), were also 
observed (Frail et al, 1997), providing a nice double-check on the physical 
dimension of the source of $\sim$ 0.1 pc.  An interesting, and less expected 
result is that the scintillation is of large amplitude and broad band, 
suggesting it is diffractive (Waxman, etal, 1997). This requires a small size, 
which comes from the fact that the intensity is concentrated in a ring of 
radial extent substantially smaller than the radius of the visible disk 
(Waxman, 1997c, Panaitescu \& \Mesz, 1997b, Sari, 1997b). This is because 
for equal observer times one sees an egg-shaped region of the outflow
and the portion around the edges corresponds to a younger, hence hotter and 
higher field, stage of the remnant. Another unexpected feature is that the 
optical light curve appears to have been steady or decaying for a brief (few 
days) initial period before it started to rise and then decay (Pedersen, 
1997). One explanation for an initial decay could be that it is due to 
emission from a central jet, which later becomes overwhelmed by emission 
from a more energetic isotropic outflow at large angles (\Mesz, et al, 1997).

One issue is whether the fireball, as it slows down by sweeping increasing
amounts of external matter, evolves with $\Gamma \propto r^{-3/2}$ as
expected in the adiabatic limit, or as $\Gamma \propto r^{-3}$ as expected 
if the remnant is in the radiative regime (Rees \& \Mesz, 1992). This would
have consequences for the evolution of the afterglow (Vietri, 1997b, Katz \& 
Piran, 1997). In the latter case, the remnant would evolve faster,  
and could reach the nonrelativistic regime sooner, even if after some time 
it becomes adiabatic, as it should.  Physically, however, for the remnant 
dynamics to be `radiative' implies that most of the kinetic energy in 
protons and fields has to be radiated in less than a dynamic time (\Mesz, 
\etal, 1997). This would require field reconnection, as well as efficient
mechanisms for protons to re-energize electrons whose cooling timescale is 
shorter than the dynamical time in the cooling region (behind the shock 
front and throughout the remnant), and it is far from being understood how
this would happen. In fact, the 
optical power law of GRB 970228 is unbroken so far, after 8 months,
arguing for an early onset of the adiabatic regime, and indicating that the
nonrelativistic regime has not been reached yet (which is strong indication
for a cosmological distance, Wijers et al, 1997). Another observational
constraint comes from the radio scintillations in 970508: this requires
a relatively small size $\lesssim$ few $10^{17}$ cm, after a time of several 
weeks. The longitudinal size is $r\sim 4(2n+1) \Gamma_o^2 c t$ where 
$n=(3/2,3)$ for an (adiabatic, radiative) remnant, but the ring structure of 
the remnant emitting region reduces somewhat the coefficient in front 
(Panaitescu \& \Mesz, 1997b). The adiabatic behavior seems more in accord 
with observations (Waxman, 1997c). However numerical calculations of the light 
curve and spectral evolution (e.g. Panaitescu \& \Mesz, 1997c) are needed
in order to address this issue more thoroughly.

A question is why some bursts (e.g.\ GRB 970111) are detected in 
$\gamma$-rays but not in X/O, despite being in in the field of view of 
Beppo-SAX.  One reason may be that the $\gamma$-ray emission could be due to 
internal shocks (leaving essentially no afterglows \cite{mr97a}) and the
environment has a very low density, so the external shock occurs at larger 
radii and over longer times than in ``canonical" afterglows, resulting in 
a sub-threshold X-ray intensity.  This may be the case if GRB arise from 
compact binaries which are ejected to considerable distances from the host 
galaxy, where the external density may be much lower than the typical ISM 
values. Low density environments may also occur if the GRB goes off inside 
a pulsar cavity inflated by one of the precursors in the binary.  This give
rise to a deceleration shock months after the GRB with a much lower brightness.
Conversely, an interesting consequence of anisotropic models (Wijers et al, 
1997; Rhoads, 1997; \Mesz, et al 1997) is that there could be a large fraction 
of detectable afterglows for which no $\gamma$-rays are detected. The outflow
at large angles is certain to be more baryon-loaded, and therefore of lower
$\Gamma$, so that the shocks would occur later and would be at longer 
wavelengths.  

It is also possible that some bursts arise in unusually high density 
environments (such as a star-forming region, where failed supernova or 
hypernova progenitors may reside (Paczy\'nski, 1997). This could lead to a more
rapid onset of the deceleration leading to the X-ray phase, and it would also
imply an increased neutral gas column density and optical depth in front of
the source.  A special case is that of GRB 970828,
where X rays have been observed, but no optical radiation down to faint
levels (Groot \etal, 1997). The presence of a significant column density
of absorbing material has been inferred from the low energy turnover of
the X-ray spectrum (Murakami, \etal, 1997), and the corresponding dust 
absorption may in fact be sufficient to cause the absence of optical emission 
(Wijers \& Paczy\'nski, private comm.). The difference between the low
density and high density environments cases could be tested if future
observations of afterglows reveal a correlation with the degree of galaxy
clustering or with individual galaxies.

\noindent
I am grateful to M.J. Rees for stimulating collaborations in this subject,
as well as to H. Papathanassiou, A. Panaitescu and R. Wijers.
This research has been supported in part by NASA NAG5-2857.

\end{document}